\documentclass[twocol]{ametsocV5}

\usepackage{amsmath,amsfonts,amssymb,bm}
\usepackage{mathptmx}%{times}
\usepackage{newtxtext}
\usepackage{newtxmath}

\title{The physics of AI weather models}

    \authors{George Craig\correspondingauthor{George Craig, george.craig@lmu.de},
    Tobias Selz \thanks{Current affiliation: Karlsruhe Institute of Technology.},
    Matthias Beylich,
    and Kirsten I. Tempest
 }

     \affiliation{Meteorological Institute, LMU Munich}

\abstract{Could it be that AI weather models are solving physical equations, although they may not be the equations used by conventional NWP models? We compute correlations of forecast skill and Centered Kernel Alignment, providing evidence that different AI weather models represent the atmosphere in similar ways, despite differences in architecture and capacity. We argue that the architecture and training of the AI models constrains the form of the physical laws that they might simulate. In particular, we propose that the models implement a particle description of the atmosphere, where the latent variables at each mesh point correspond to the position of a particle in the high dimensional latent space. We hypothesize that the movement of the particles follows a gradient flow in the latent space towards a minimum of a learned free energy functional. Analysis of the GraphCast and Aurora models show that they make changes on large spatial scales in the early processor layers and move to smaller scale with increasing layer depth, consistent with the gradient flow hypothesis.} 

\begin{document}

%% Necessary!
\maketitle

%%%%%%%%%%%%%%%%%%%%%%%%%%%%%%%%%%%%%%%%%%%%%%%%%%%%%%%%%%%%%%%%%%%%%
% SIGNIFICANCE STATEMENT/CAPSULE SUMMARY
%%%%%%%%%%%%%%%%%%%%%%%%%%%%%%%%%%%%%%%%%%%%%%%%%%%%%%%%%%%%%%%%%%%%%
%
% If you are including an optional significance statement for a journal article or a required capsule summary for BAMS 
% (see www.ametsoc.org/ams/index.cfm/publications/authors/journal-and-bams-authors/formatting-and-manuscript-components for details), 
% please apply the necessary command as shown below:
%
% \statement
% Significance statement here.
%
% \capsule
% Capsule summary here.

%%%%%%%%%%%%%%%%%%%%%%%%%%%%%%%%%%%%%%%%%%%%%%%%%%%%%%%%%%%%%%%%%%%%%
% MAIN BODY OF PAPER
%%%%%%%%%%%%%%%%%%%%%%%%%%%%%%%%%%%%%%%%%%%%%%%%%%%%%%%%%%%%%%%%%%%%%
%

\section{AI weather models and physical laws}

Artificial intelligence (AI) models have recently matched and even surpassed the forecast skill of conventional Numerical Weather Prediction (NWP) models on a variety of scores \citep{biAccurateMediumrangeGlobal2023,lamLearningSkillfulMediumrange2023,pathakFourCastNetGlobalDatadriven2022,bodnarFoundationModelEarth2025}. AI models are providing daily weather forecasts \citep{ecmwfECMWFCharts2025}, and operational weather services, as well as technology companies, are investing heavily in developing new AI models.

While the skill of the new AI models is acknowledged by atmospheric scientists, these models are data-driven and frequently criticized as ``black boxes". They produce predictions, but do not enable physical understanding. Internally AI models use latent variables with no direct interpretation as physical quantities, go through many stages of nonlinear calculations, and rely on many (in some cases over a billion) learned parameters. It is possible to include physical laws in AI models, e.g. Physics Informed Neural Networks \citep{cuomoScientificMachineLearning2022}, but this is not what the recent generations of AI weather models do. The philosophy is reminiscent of David Mermin's famous caricature of the Copenhagen interpretation of quantum mechanics \citep{merminWhatsWrongThis1990}: ``Shut up and calculate!"

This situation is not very satisfying to a scientist. Indeed, the very next sentence in Mermin's essay is ``But I won't shut up." We are curious; we use models to help us explore and understand the complexity of the physical world. Physical understanding is the basis for determining where and how models can be applied. This is particularly critical in the case of data-driven models, where it is not \textit{a priori} obvious that they can produce valid predictions of situations that are not strictly within the bounds of their training data. Recent models have shown good performance in predicting daily weather patterns, for quantities that they have been trained on, even though the weather never exactly repeats. However, they often fail for quantities that have not been included in the training.

Many of these issues have been improved by further developments to the models or their training regimes, but other deficiencies such as the inability to produce rapid error growth in response to small initial perturbations \citep{lorenzPredictabilityFlowWhich1969,palmerRealButterflyEffect2014,selzCanArtificialIntelligenceBased2023} may indicate more fundamental limitations of the data-driven approach. This concern is crucial for the application to climate change: can a data-driven model make accurate projections of a climate that has never occurred? And how can we assess this if we cannot trace critical aspects of the projection back to the inputs and assumptions that are responsible? Furthermore, these projections are the basis for scenarios and storylines that are used by decision-makers \citep{shepherdStorylinesAlternativeApproach2018}, and require consistency between all aspects of the model output, including quantities that are not observed and cannot be directly learned or evaluated \citep{olivettiDatadrivenModelsBeat2024,sunCanAIWeather2025}.

It is clear that physical understanding of data-driven atmospheric models would be desirable, but it is not clear what form this understanding would take. The pessimistic view is that the models do a sophisticated form of curve fitting and interpolation, but there is no physics learned and no expectation of generalization to problems where the performance has not been directly verified. A more optimistic view is that the models implement the same underlying physical laws as current NWP models but perhaps in an obfuscated form, and with extra tuning of the outputs to closer match observations. If this view could be demonstrated, then much current knowledge of the atmosphere and atmospheric modeling could be applied to the new models. A third, more interesting, possibility is that the new models encode a new and different physical understanding of how the atmosphere works, in which case they could be a source of new insights.

Our task is then to take an AI weather model, as implemented in code and trained parameters, and to extract physical principles. This is not easy -- imagine being given the code of a modern NWP or climate model and being asked to write down the equations that it implements. With some fore-knowledge, one could probably identify the dynamical core and find the numerically approximated fluid equations, but one would also find parameterizations, that are based on additional physics (e.g. radiative transfer) or more abstract representations of the underlying dynamics (e.g. turbulence theory or mass flux convection). The model contains parameters obtained by fit to observations or laboratory experiments, and finally, some values that are simply tuned to produce the desired results.

\begin{figure}[h]
\centerline{\includegraphics[width=19pc]{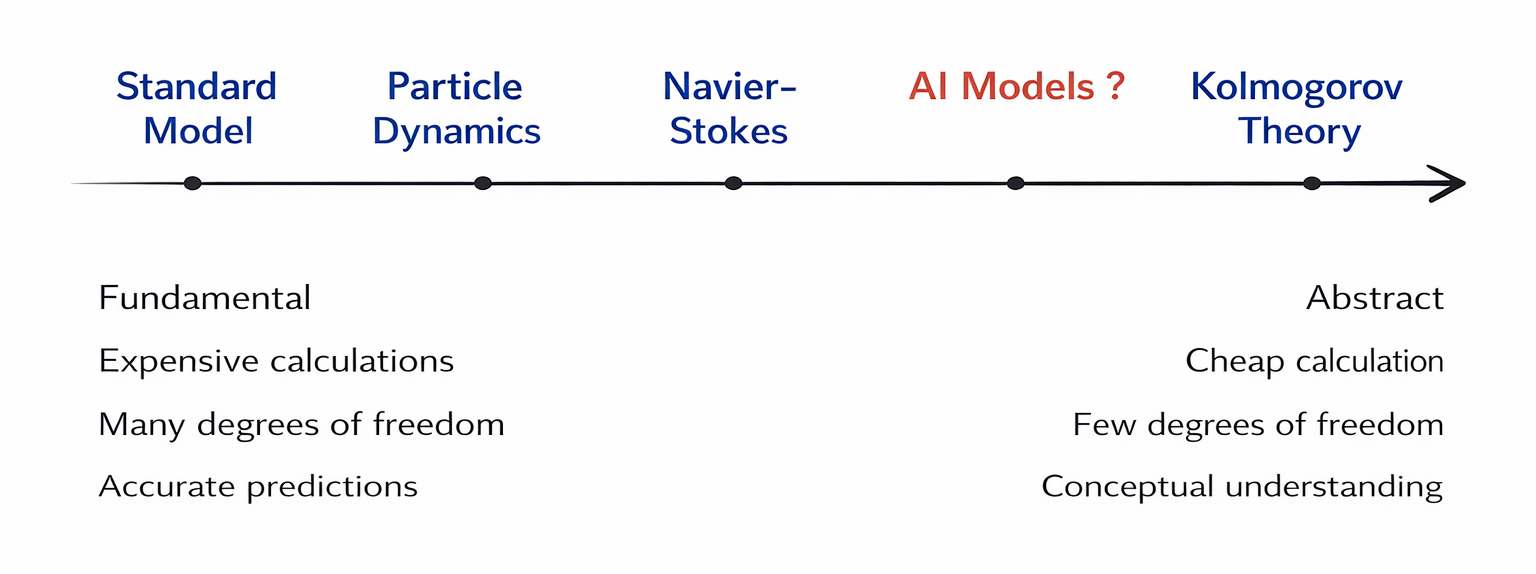}}
 \caption{The hierarchy of physical theories.}\label{f1}
\end{figure}

So how will we recognize fundamental physical laws when they are expressed in an unfamiliar and approximated form? In atmospheric modeling, the fundamental representation is usually considered to be the equations of fluid motion, Navier-Stokes or some approximation, but this is not the only option. As illustrated in Fig.~\ref{f1}, there is a hierarchy of physical theories. The fluid equations are an approximation to molecular motions, but one could go to even more fundamental descriptions of nature, and eventually all the way to the standard model of particle physics. 

In general, as we move to a more fundamental level of description, we gain accuracy and generality but lose computational efficiency. On the other hand, as we move up the hierarchy, the physical models are  based on entities that are emergent phenomena in lower level solutions. For example, under the right conditions, molecular motions will organize into hydrodynamic streaming flows that can be represented efficiently in a continuum approximation. The number of degrees of freedom is reduced and the computation of solutions becomes more efficient, but the range of validity of the higher level equations is only a subset of the situations covered by the more fundamental representation. 
These trends continue as we move to more abstract theories of fluid motions, such as shell models of turbulence, or even the Kolmogorov scaling for the kinetic energy spectrum of homogeneous and isotropic turbulence. The motivation for these simplifications is not computational efficiency but rather the more comprehensive understanding that is possible in a simplified model. 

The aim of this paper is to advance the hypothesis that AI weather models are indeed representing physics, but not as fluid equations, rather as a new type of theory that is more abstract. If this is true, there is potential not only for models that run faster, but that also give new physical insights into the atmospheric flow.

Our specific contributions are:
\begin{itemize}
    \item empirical evidence from correlations of forecast skill and CKNNA of latent space states that suggests that AI weather models with diverse architectures have similar underlying representations of the atmosphere. (Section~\ref{sec:corr})
    \item analysis of model architectures, taking advantage of recent mathematical results on continuum limits of neural networks, to propose that the different models are discretizations of a common set of equations, describing a system of interacting particles in (thermodynamic) equilibrium. (Sections~\ref{sec:constraints}--\ref{sec:kinematics})
    \item a proposal that the training of the networks favors a particle dynamics in the form of a gradient flow that minimizes a free energy. (Section~\ref{sec:dynamics})
    \item discussion of observable consequences of the hypothesized equations, consisting of the proposal that gradient flow implies that early processor layers should affect larger spatial scales, which is confirmed for the GraphCast and Aurora models, and some suggestions for how methods from the mechanistic interpretability literature could be applied. (Section~\ref{sec:discussion})
\end{itemize}

\section{Are all AI weather models alike?}
\label{sec:corr}

If AI weather models are a new representation of physical laws, they should in some sense behave similarly, despite differences in architecture and scale (number of parameters). In this section, we consider two measures of similarity, one based on forecast skill (similar behavior) and one based on internal representation (similar learned structure).

\subsection{Similarity of model performance}

\begin{figure}[h] 
\centerline{\includegraphics[width=19pc]{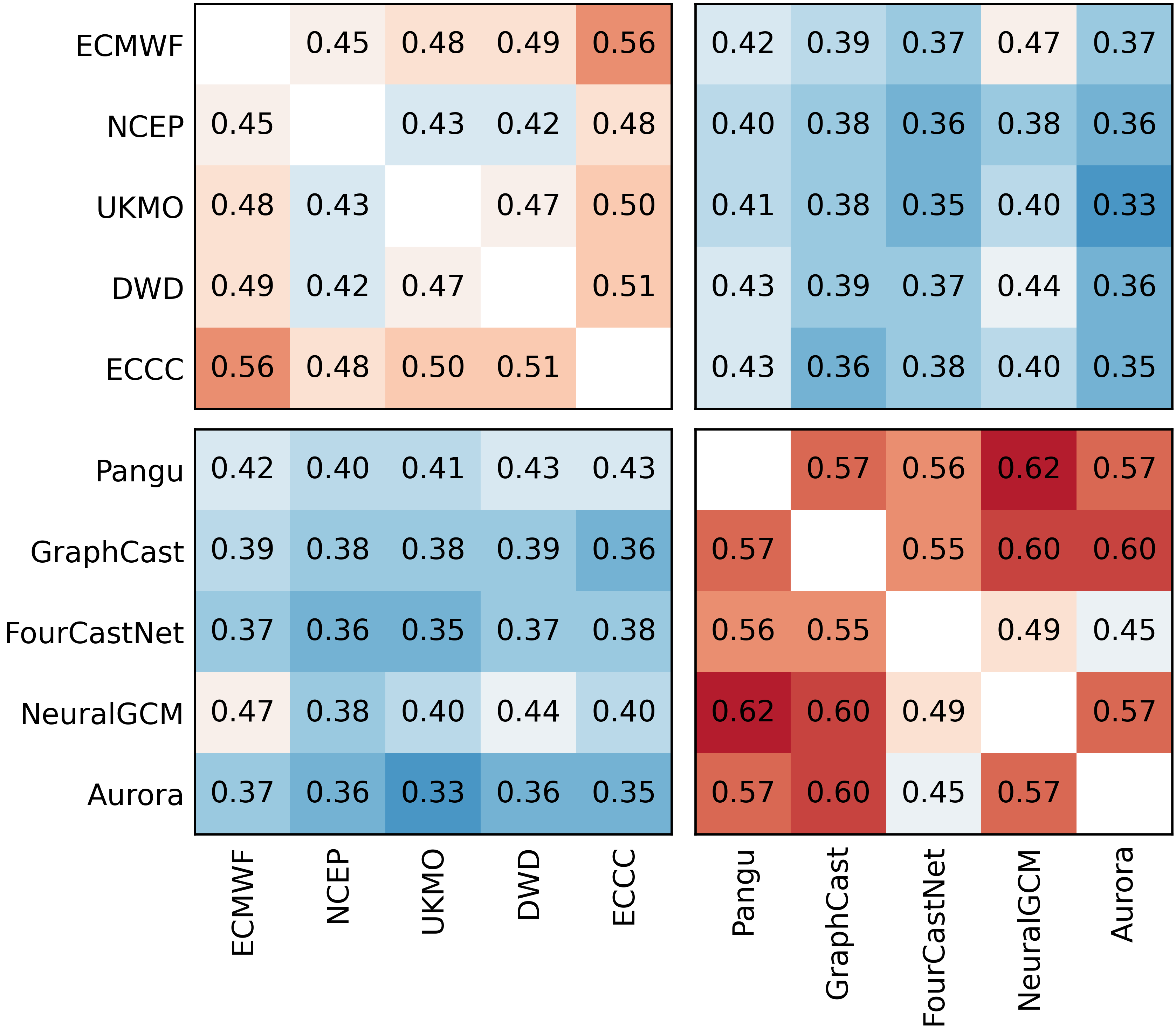}}
 \caption{Correlation matrix of RMSE timeseries for several NWP (upper group) and AI weather models (lower group). The root-mean-square error (RMSE) is computed from global 5-day forecasts of the 500~hPa geopotential against ERA5 for the period 2021--2025. Forecasts are started every 12 hours.}\label{fcorrmatrix}
\end{figure}

The initial motivation for the hypothesis that AI weather models are a different representation of physics from NWP models came from examining time series of forecast skill for a large number of different models. 
The plot shows that the AI models correlate more strongly among themselves than with the NWP models, that is, they tend to be good on the same days and poor on the same days (see Appendix \ref{app:corr} for more details).
We note that a confounding factor in this analysis is that the AI models all start from ERA5 analyses, whereas the weather centers operating different NWP systems have their own initial conditions. As discussed in the appendix, this increases the correlation among the AI models at early lead times, but does not appear to affect the correlations of 5~day forecasts, as plotted in the figure.
This is consistent with the hypothesis that the AI weather models have something in common that differs from the conventional models, but does not give much insight into what the common elements might be.

\subsection{Universality of the latent space representation}

The correlation of forecast skill between different AI weather models raises the possibility that the models themselves are in some sense similar. Now we will consider whether the internal representations of atmospheric states learned by different models is also similar. It is not straight-forward to compare the internal structure of AI models since the meteorological information is encoded into latent space variables that have no direct physical interpretation. Even the dimension of the latent space varies between models.

To compare the latent space representations of different models, we will take advantage of the fact that we can compare different mesh points within each model individually. Following work in other applications of machine learning \citep{huhPlatonicRepresentationHypothesis2024,edamadakaUniversallyConvergingRepresentations2025}, we will rephrase the question to ask whether the two models consider the same pairs of mesh points to be similar. In particular, for each model, we compute the inner product between latent space vectors of pairs of mesh points corresponding to different times. The inner product computed from one model can be compared to that of the other model for pairs of mesh points corresponding to the same locations and times. The normalized correlation between the values computed by the two models provides a measure of the extent to which the two models are representing similar states by similar vectors. Formally, this method is referred to as Centered Kernel Analysis (CKA), and is the normalized version of the Hilbert-Schmidt independence criterion \citep{kornblithSimilarityNeuralNetwork2019,huhPlatonicRepresentationHypothesis2024}.

A potential problem with CKA is that the variation in inner-product similarity may be dominated by pairs of points at different times of year, so that a high correlation between different models could arise simply from the seasonal cycle. To provide a more specific test, we will also consider a variant of CKA, Centered Kernel Nearest Neighbor Alignment (CKNNA), which restricts the correlation to the subset of mesh point pairs that are most similar.

CKA and CKNNA take the form of correlation coefficients, where a value of 1 indicates perfect correlation and a value of 0 shows no correlation at all. \citet{edamadakaUniversallyConvergingRepresentations2025} computed CKNNA among 59 different molecular structure models, and obtained values ranging from 0.03 and 0.55 for models with different architectures and up to 0.85 for models with similar architectures. Even more ambitiously, \citet{huhPlatonicRepresentationHypothesis2024} compared language and vision models and obtained CKNNA values between 0.06 and 0.22. While these values are not particularly large, they are significantly different from zero, and suggest that even models trained on different data modalities may have similarities in their representations of the world. In both studies, larger models tended to correlate better, suggesting that there may be some convergence to a common representation as the expressivity of the models improves.

As a first examination of AI weather models we will compute CKA and CKNNA between two models with very different architectures and sizes: GraphCast, a graph neural network with 37 million parameters \citep{lamLearningSkillfulMediumrange2023}, and Aurora, a transformer model with 1.1 billion parameters \citep{bodnarFoundationModelEarth2025}. We will compare the latent states of both models after the last processor layer, just before the decoder. To avoid latitudinal variability dominating the correlations, correlations will be computed for single geographical locations. Results will be shown for CKA and for CKNNA with 40 nearest neighbors. See Appendix \ref{app:cka} for details of the calculations. 

\begin{figure}[h]
\centerline{\includegraphics[width=19pc]{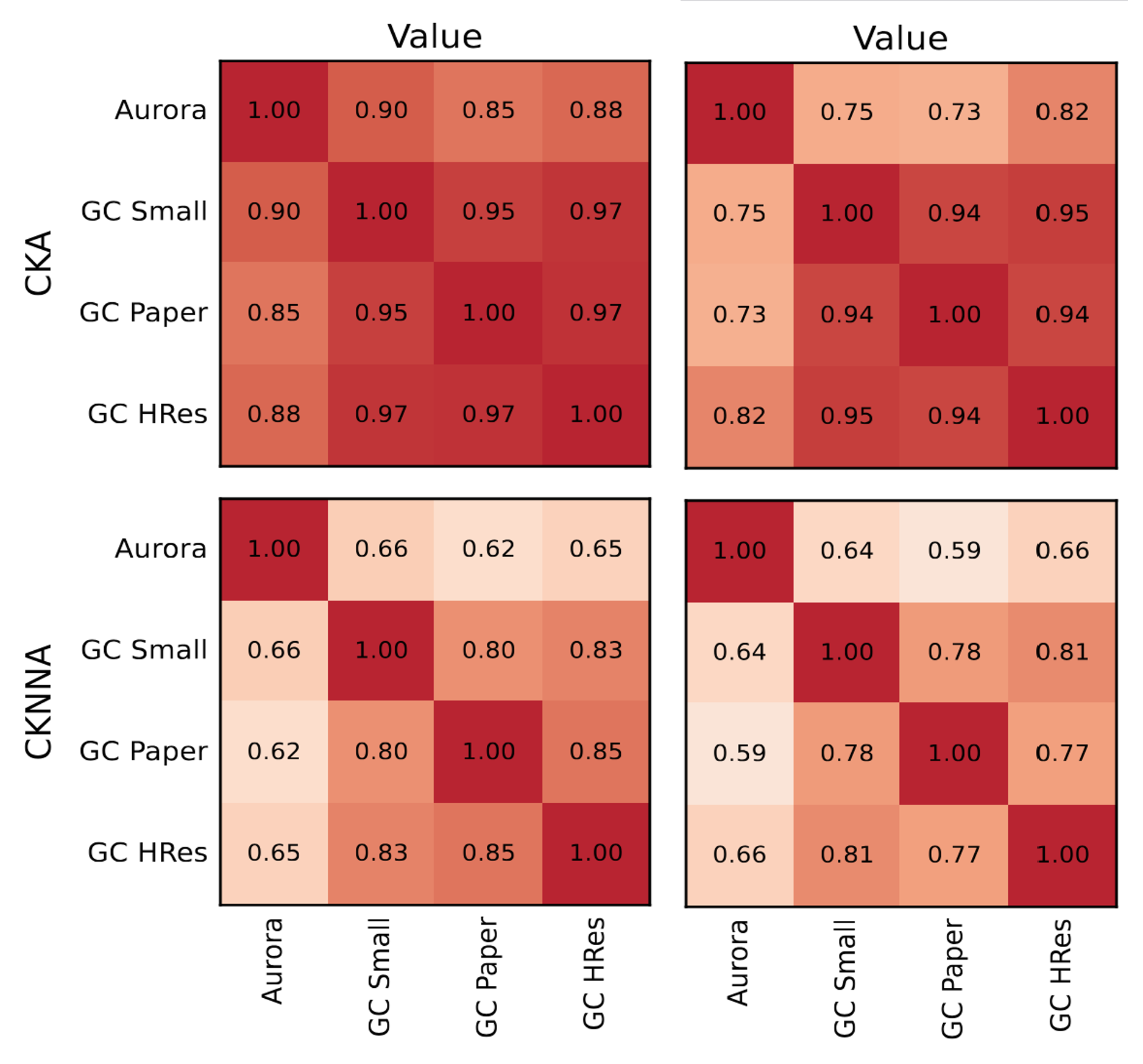}}
 \caption{Centered Kernel Alignment (top row) and Centered Kernel Nearest Neighbor Alignment (bottom row) between Aurora and three versions of GraphCast. First column for a midlatitude point (50°N, 9°W), and second column for a tropical point (1°S, 0°E).}\label{f6}
\end{figure}

A typical example of the results is shown in Fig.~\ref{f6}. Correlations were computed for a number of different locations, but the figure shows only results for data spanning the whole of 2020 at two locations. Results for other locations were very similar. The correlation matrices show very clearly that the three versions of GraphCast are highly aligned with each other, with CKA and CKNNA values always greater than 0.94 and 0.77 respectively. Comparing GraphCast and Aurora, the values are slightly lower, with values always greater than 0.73 and 0.59 for the CKA and CKNNA respectively. There appears to be a significant alignment between the latent state representations of GraphCast and Aurora, despite the differences in architecture and size.

\section{Constraints from model architecture and training}
\label{sec:constraints}

In the following discussion, we will focus on global weather prediction models designed to make deterministic weather forecasts for periods out to about two weeks. Examples of such models include Pangu \citep{biAccurateMediumrangeGlobal2023}, GraphCast \citep{lamLearningSkillfulMediumrange2023}, FourCastNet \citep{kurthFourCastNetAcceleratingGlobal2023}, AIFS \citep{langAIFSECMWFsDatadriven2024}, Aurora \citep{bodnarFoundationModelEarth2025}, and ArchesWeather \citep{couaironArchesWeatherEfficientAI2024}. Since the architectures are similar, it is possible that the discussion will also apply to a single realization of a model trained to produce ensemble members that sample a forecast distribution, such as AIFS-CRPS \citep{langAIFSCRPSEnsembleForecasting2024}, or a single step of the denoising process in generative diffusion or flow-matching models including GenCast \citep{priceGenCastDiffusionbasedEnsemble2024}, ArchesWeatherGen \citep{couaironArchesWeatherGenSkillfulComputeefficient2026}, and FGN \citep{aletSkillfulJointProbabilistic2025}. This possibility will be considered in more detail in Section~\ref{sec:discussion}.

While there is much diversity among the class of models we are considering, there are also some significant common elements. These include aspects of the problem formulation, the internal representation of the atmospheric state, the architecture of the model, and the training. These will be reviewed in the following subsections, and the implications for the physical representation will be explored in the following sections.

\subsection{Problem formulation}

The calculation is posed as a mapping from an initial state $X_I$, to a forecast state $X_{F}$ at a later time. The initial state is obtained from an analysis data set such as ERA5, and consists of a selection of meteorological variables defined on a grid, along with some auxiliary information such as time of day. Often, two previous timesteps are included in the initial state vector.
The AI model is trained to minimize the expectation of a loss function over a large training data set. The loss function expresses the difference between the forecast state $X_{F}$ and the analysis at the forecast time $X_{A}$.
Schematically, it takes the form:
\begin{equation}
      L=\parallel X_{F} - X_{A} \parallel ^r
\end{equation}
where $r=$ 1 or 2. This simplified form of the loss function implicitly includes any normalization and weighting of the different meteorological variables, as well as a possible sum of terms over a roll-out training interval. Note that the choice between these two norms does not seem to influence the behavior of the resulting model significantly, as noted by \citet{selzEffectiveResolutionAI2025}.

\subsection{Latent state}

\begin{figure}[h]
\centerline{\includegraphics[width=19pc]{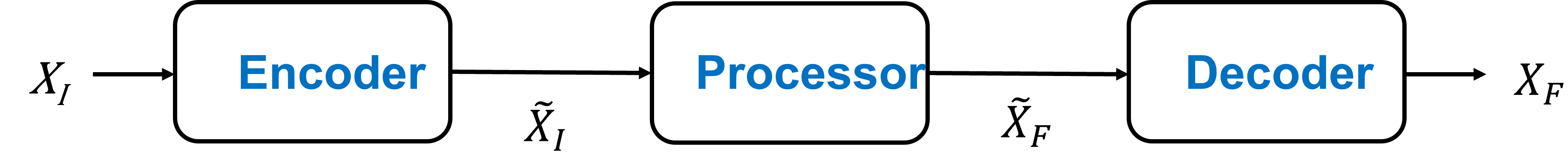}}
 \caption{The encoder-processor-decoder architecture.}\label{f2}
\end{figure}

The models are constructed with an encoder-processor-decoder architecture (Fig.~\ref{f2}). The encoder is typically a learned (nonlinear, e.g. Multi-Layer Perceptron (MLP)) transform that maps the meteorological variables to an internal representation. The internal representation $\tilde{X}_I$ is defined on a spatial mesh of $M$ nodes, where the state at each mesh point $p$ is defined by a vector $\tilde{x}_p$ in a latent space of high dimension, e.g. $D=512$, for a total of $M\times D$ values. The encoder aggregates information, combining neighborhoods in the horizontal and vertical directions (usually, but not always, complete columns), to compute the latent state vector at each internal mesh node. The processor now operates through a series of layers that communicate between mesh nodes and apply nonlinear transformations to produce the final prediction in the latent space, $\tilde{X}_F$. A decoder is then applied to transform the mesh node values in the latent space back to meteorological variables on a conventional grid. The encoder and decoder are both learned during training, and are not generally inverses of each other.

\subsection{Processor blocks}

\begin{figure}[h]
\centerline{\includegraphics[width=19pc]{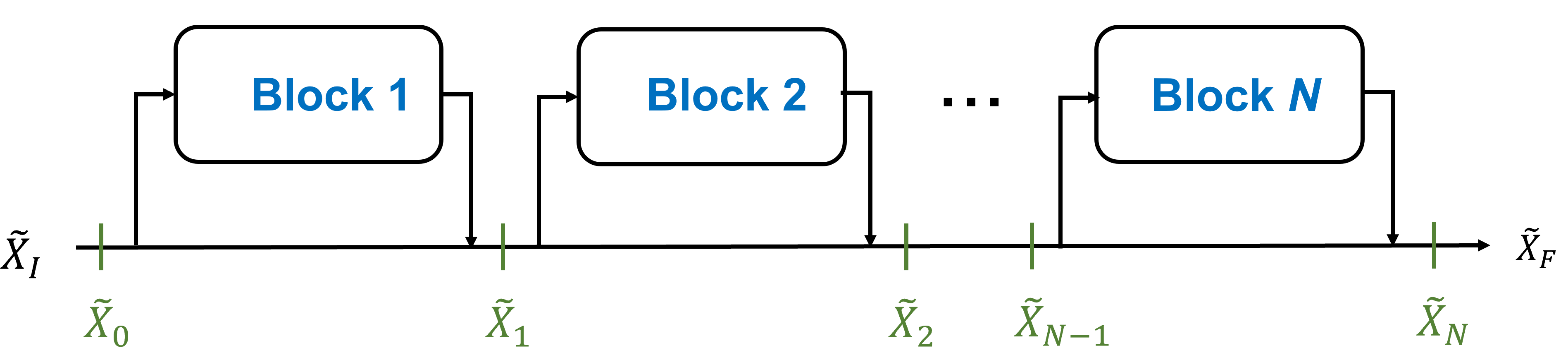}}
 \caption{The data flow in a residual network architecture.}\label{f3}
\end{figure}

The processor follows a residual network architecture, with a succession of $N$ processor blocks whose output is summed to a residual connection that bypasses the block. The processor can thus be seen as a series of blocks that read from, and write to, a residual stream that passes from encoder to decoder \citep{elhageMathematicalFrameworkTransformer2021}, as shown in Figure~\ref{f3}. Each block, $B_i$ for $i=1\cdots N$, adds an increment to the latent space vector 
    \begin{equation}
    \label{eq:resnet}
        \tilde{X}_i = \tilde{X}_{i-1} + \frac{1}{N} B_i(\tilde{X}_{i-1}) ,
    \end{equation}
where $\tilde{X}_0 = \tilde{X}_I$ and $\tilde{X}_N = \tilde{X}_F$ are the initial (post-encoder) and final (pre-decoder) states, resp., in the latent space.
The constant $1/N$ is is usually absorbed in the processor weights, but has been factored out of the function $B_i$ here so that the sum of the increments does not diverge as the number of layers becomes very large. This limit will be considered in the next section.

Finally, each block is composed of two stages: an interaction stage, where different mesh points influence each other, followed by a MLP stage, that processes the latent state at each mesh point individually. In a transformer architecture, the interaction stage takes the form of a self-attention layer, while in a graph neural network it consists of an edge update followed by a concatenation of the sum of the edge values arriving at each mesh node onto the latent space vector. Usually a layer-norm is applied to the input (pre-norm) or the output (post-norm) of the block. This controls the effective magnitude of the latent space vector.

\subsection{Training and regularization}
\label{sec:training}
The training of an AI weather model seeks to minimize a loss function that quantifies the error in the model forecast. However, the loss function may have many minima, and the actual weights that result are strongly influenced by constraints applied during the training, both explicit and implicit. These usually take the form of regularization constraints that are intended to make the trained model more robust to small changes in the data.
Modern models seldom apply a simple explicit regularization like a weight decay term in the loss function, but similar effects can be achieved implicitly with stochastic gradient descent and AdamW . These methods penalize large magnitudes of weights, biasing the model towards simpler and smoother functions \citep{smithBayesianPerspectiveGeneralization2018,loshchilovDecoupledWeightDecay2019}. Another regularization method that has been recently applied in the Aurora model \citep{bodnarFoundationModelEarth2025} is drop path, stochastically removing entire processor layers during training \citep{huangDeepNetworksStochastic2016}. Rather than penalizing large weights, this penalizes large processor increments, directly influencing the model dynamics. It is important to bear in mind, however, that unlike the model architecture, regularization will only bias the model towards certain dynamics, rather than applying a hard constraint.

\section{Kinematics}
\label{sec:kinematics}

\subsection{The ODE limit}

The residual network structure seen in Eq.~\ref{eq:resnet} suggests that dividing the computation into an increasing number of blocks should lead to an ordinary differential equation that flows from the initial to final state in the latent space. In particular, we define a coordinate $s_i = i/N$, which measures progress through the network, with each block advancing through an interval $\Delta s = 1/N$,
    \begin{equation}
    \label{eq:resnetode}
        \tilde{X}_i = \tilde{X}_{i-1} + \Delta sB_i(\tilde{X}_{i-1}), \quad i = 1 \ldots N.
    \end{equation}  
This leads to 
     \begin{equation}
    \label{eq:ode}
        \frac{d\tilde{X}(s)}{ds} = B(\tilde{X}(s),s), 
    \end{equation}  
as $N \to \infty$. The variable $s$ measures progress through the processor blocks, and $B(\cdot,s)$ is a smooth function of $s$. The correspondence between differential equations and limits of neural networks has been studied under the name of Neural ODEs, and has been shown to hold rigorously for a number of architectures \citep{chenNeuralOrdinaryDifferential2018}. In particular, it holds for some deep ResNets where the processor blocks are standard Multi-Layer Perceptrons \citep{marionImplicitRegularizationDeep2024}. Note that convergence proofs will generally require some regularity assumptions on the processor increments to ensure that the limit exists.

\subsection{Symmetries and the particle dynamics description}
A system of ordinary differential equations defines a dynamical system. The equations represented by a neural network are very general, but there are some limitations imposed by the network architecture. In this section, we will examine the symmetries of the interaction and MLP blocks, to see how they constrain the dynamics that can be represented.

The interaction block in the processor determines how the different mesh points influence each other. Different AI weather models take very different approaches. A graph neural network like GraphCast \citep{lamLearningSkillfulMediumrange2023} or FastNet \citep{dunstanFastNetImprovingPhysical2025} defines a hierarchy of connections. Mesh nodes are linked to their nearest neighbors, and selected mesh nodes have links to other nodes further away. The connections are chosen to become increasingly sparse with distance.
Transformer models such as PanguWeather \citep{biAccurateMediumrangeGlobal2023} or Aurora \citep{bodnarFoundationModelEarth2025} are fully connected within window regions that each cover part of the domain. By shifting the window positions between processor layers, it becomes possible for mesh nodes in different windows to communicate.
With either architecture, information can cross between any two locations on the globe within a few processor layers.

In general it is not possible to connect every mesh node to every other because the number of connections grows quadratically with the number of nodes. For practical computations, AI weather models reduce the number of connections, but it doesn't seem to be very important how it is done. The pattern of connections is defined without regard to the weather or geography, and the different architectural choices do not seem to have a critical effect on the performance of the model. They appear to be made for computational efficiency rather than reflecting assumptions about the underlying physical processes. 
Motivated by these considerations, we hypothesize that the underlying equations represented by the AI weather models are in fact fully connected; every location can in principle influence every other location.

Now consider the MLP block. An important symmetry that is present in all of the models discussed here comes from using shared weights for all mesh nodes. Rather than hard-coding the location of a node, it is included in the latent state vector via a positional encoding or bias, along with other position-dependent information like orgraphic height or land-sea mask. The same transformation is applied to all mesh points by the MLP, but the results are different, since the location is included in the input vector. If the mesh points were fully connected (as hypothesized above), the models would be invariant to relabeling of the mesh points, since they carry all information about their location as part of their state.

This permutation equivariance suggests that we regard the system of mesh points as a set of interacting particles, where each particle has a vector of properties, including its geographical location \citep{vuckovicMathematicalTheoryAttention2020,sanderSinkformersTransformersDoubly2022,geshkovskiMathematicalPerspectiveTransformers2023}. 
The particles move through configuration space (specified by latent variables) with each processor block updating their position in that space. In the ODE limit defined by Eq.~(\ref{eq:ode}), the processor defines the rate of change of the latent space vector, that is, the velocity in configuration space.

The interaction part of the processor contributes to the velocity with the sum of the influences of other particles, while the MLP contribution depends on the state of the individual particle. In transformer models, each contribution is added to the residual stream in turn, giving a form of operator splitting. In graph network models, the interactions are computed by first updating a set of edge nodes corresponding to each connection between two mesh nodes, then summing the contributions of all the edge nodes that are directed towards a given mesh node. This sum is concatenated to the mesh node state to form the input to the MLP, giving a different form of operator splitting. While it has not been rigorously proven, it seems plausible that the two forms of splitting lead to similar equations for a fully connected network in the ODE limit.

\subsection{The mean-field limit}
The view of an AI model as a particle system has been recently reviewed by \citet{peyreOptimalDiffusionTransports2025} and \citet{rigolletMeanFieldDynamicsTransformers2026}. An important consequence of this view is existence of a mean-field limit description in the limit of an infinite number of particles. 
Following \citet{peyreOptimalDiffusionTransports2025}, but adapting the notation for consistency, we see that the system of ODEs in Eq.~(\ref{eq:ode}) defines the evolution of a group of interacting particles. The particles move through the latent space, although their physical coordinates are fixed and encoded as part of the latent space vector.
Each particle can then be regarded as a sample from a continuous distribution $\tilde{\alpha}_s$, the particle density, i.e. the number of particles per unit of volume in the latent space. 

Choosing a set of mesh points in physical space is equivalent to approximating $\tilde{\alpha}_s$ with the empirical measure $\tilde{\alpha}_s \approx \frac{1}{M} \Sigma_p \delta(\tilde{x}-\tilde{x}_p)$, where $\tilde{x}_p$ is the latent space vector at mesh point $p$ and $M$ is the number of mesh points. 
In this view, a high resolution model has many particles, and can be expected to give more accurate results because it better samples the density function, $\tilde{\alpha}_s$.

To preserve probability density, $\tilde{\alpha}_s$ must satisfy a continuity equation in the latent space
      \begin{equation}
      \label{eq:continuity}
        \partial_s \tilde{\alpha}_s + \nabla \cdot (\tilde{\alpha}_s v_s) = 0,
    \end{equation} 
where the divergence is taken with respect to the latent space variables. The latent space velocity $v_s$ measures the rate of change of latent space position with progress through the processor layers, and is defined by
\begin{equation}
%    \label{eq:velocity}
    v_s (\tilde{x}_p) \equiv B_p(\tilde{X}(s),s) = \frac{d\tilde{x}_p}{ds} ,
\end{equation} 
where $B_p(\tilde{X}(s),s)$ denotes the processor increment applied at mesh point $p$.
The problem is now to characterize the learned velocity field $v_s$.

\section{Dynamics}
\label{sec:dynamics}

The previous two sections argued that if AI weather models indeed represent discretizations of differential equations, the architecture constrains the forms that these equations might take. But this has not given a closed set of equations since the form of the velocity is specified by the model weights that are learned as the model is trained. The loss function compares the output of the network with the training data, but does not directly constrain the velocity. Many different paths through the configuration space could lead to the same end point. However, the velocity is not completely free since the weights at each stage of the processor are influenced by the regularization. 

As noted in Section~\ref{sec:constraints}\ref{sec:training}, regularization methods such as weight decay and drop path will bias the model towards uniform increments from the different processor blocks for each mesh point. In the language of particle dynamics, this is a bias towards uniform velocity. The system of particles is pressured to flow in straight lines from the initial to the final configuration. However, a uniform flow from initial condition to endpoint is not necessarily realizable because of the need to satisfy the continuity equation (\ref{eq:continuity}), equivalently, to preserve probability mass. In this section we consider what sort of dynamics would be favored in this situation, and develop a plausible formulation of the dynamics which we term the ``Gradient Flow Hypothesis". We present first an analogy to the near-equilibrium thermodynamics of a system of particles, then show how the argument might be generalized using some relevant results from optimal transport theory. In a later section we will propose some observable consequences of the gradient flow hypothesis, which we will compare with diagnostics from some AI weather models.

\subsection{Thermodynamics of a large system of particles}

Thermodynamic equilibrium of a large system of particles is characterized by a minimum of a free energy, $\mathcal{F}$, whose form varies depending on the constraints applied during the process. In particular, the Gibbs free energy applies to systems that constrain particle number, or conserve density in the thermodynamic limit. If a system is out of equilibrium it will evolve down the gradient of the free energy towards the minimum, where the gradient vanishes. When solving for the minimum, the constraint of fixed particle number gives rise to a Lagrange multiplier $\mu = \frac{\delta \mathcal{F}}{\delta \rho}$, the chemical potential. The resulting flow is a mass flux proportional to the gradient of chemical potential $\rho v = -M_c\nabla \mu$, where $M_c$ is the mobility.

\subsection{Optimal transport}

A more general, and mathematically rigorous, version of this argument is given by optimal transport theory, which treats flows that connect two distributions of mass, minimizing a cost function while conserving mass.
Minimizing kinetic energy or equivalently distance traveled, while conserving particle number, or probability density $\tilde{\alpha}_s$ in the mean-field limit, is the classical optimal transport problem, and the solution can be expressed as a Wasserstein Gradient Flow \citep{santambrogioEuclideanMetricWasserstein2016,figalliIntroductionOptimalTransport2024}.

This simple minimization is too idealized to describe AI weather models however. The velocity of a particle is influenced by other particles and by MLP weights, so the theory needs to accommodate these influences. If these additional influences can be regarded as conservative forces, the Wasserstein gradient flow can be generalized in the so-called JKO formalism \citep{jordanVariationalFormulationFokkerPlanck1998}. Consider a distribution of mass $\alpha$, which is a function of coordinates $x$ (we will specialize to the atmospheric model $\tilde{\alpha}_s(\tilde{x})$ below). The optimal flow in this case minimizes a free energy which can take the following general form \citep{figalliIntroductionOptimalTransport2024,alvarez-melisOptimizingFunctionalsSpace2021}
\begin{equation}
    \label{eq:free-energy}
    \begin{split}
    \mathcal{G}(\alpha) = & \int H(\alpha(x))dx + \int V(x)\alpha(x)dx \\
    & + \frac{1}{2}\int\int W(x,x')\alpha(x)\alpha(x')dxdx'.
\end{split}
\end{equation} 
This free energy is the sum of an internal energy $H$, a potential energy $V$, and an interaction energy $W$.
This leads to a flow $v_s$ that satisfies the continuity equation~(\ref{eq:continuity}), with a velocity given by the Wasserstein gradient of the free energy
\begin{equation}
    \label{eq:fe-v}
    v_s = -\nabla \left( \frac{\delta\mathcal{G}}{\delta\alpha} \right) = -\nabla \left( H'(\alpha) + V + W*\alpha \right) .
\end{equation} 
Note that the variational derivative $\frac{\delta}{ \delta \alpha} \int H(\alpha(x))dx$ is equal to the ordinary derivative $H'$ since the internal energy is a scalar function of $\alpha$.
Here, the term $\frac{\delta\mathcal{G}}{\delta\alpha}$ plays the role of the chemical potential in the thermodynamic example, where the mobility is equal to $\alpha$. This form of mobility is a natural outcome since it enforces that there can be no mass flow where the density is zero.

\subsection{The gradient flow hypothesis}
\label{sec:application}

Taking inspiration from the optimal transport formulation, we hypothesize that all AI weather models represent discretizations of an equation set of the following form. The state evolves according to a flow map
\begin{equation}
    \label{eq:flow-ode}
    \frac{d\tilde{x}(s)}{ds} = v_s, 
\end{equation}  
which generates samples from a distribution that satisfies a continuity equation
\begin{equation}
    \label{eq:flow-continuity}
    \partial_s \tilde{\alpha}_s + \nabla \cdot (\tilde{\alpha}_s v_s) = 0.
\end{equation} 
We hypothesize that the velocity $v_s$ is given by the Wasserstein gradient of a free energy $\mathcal{G}$,
\begin{equation}
    \label{eq:flow-gradient}
    v_s = -\nabla \left( \frac{\delta\mathcal{G}}{\delta\alpha} \right),
\end{equation} 
or equivalently, the gradient with respect to latent space coordinates of a potential $U$:
\begin{equation}
    \label{eq:flow-potential}
    U(\tilde{x}_s) \equiv \frac{\delta\mathcal{G}}{\delta\alpha} = W*\tilde{\alpha}_s + V(\tilde{x}_s) .
\end{equation} 
The term $H'(\alpha)$ in Eq.~(\ref{eq:fe-v}) would represent an entropy term in a stochastic particle model \citep{figalliIntroductionOptimalTransport2024}, but is not present for the deterministic AI weather models considered here.

We hypothesize that the AI weather model implements a discrete form of the potential, 
obtained by approximating $\tilde{\alpha}_s$ with the empirical measure $\tilde{\alpha}_s \approx \frac{1}{M} \Sigma_p \delta(\tilde{x}-\tilde{x}_p)$. The discrete form is 
\begin{equation}
    \label{eq:flow-potential-discrete}
    U(\tilde{x}_p) \approx \sum_q W(\tilde{x}_p, \tilde{x}_q) + V (\tilde{x}_p),
\end{equation} 
where $\tilde{x}_p$ and $\tilde{x}_q$ represent the latent space vector at mesh points $p$ and $q$, respectively.
The first term is an interaction term (the attention or edge update block), connecting particles at two different locations in the latent space, while the second is a potential energy (the MLP block), that acts on each particle independently.

\subsection{Conditions for gradient flow in simplified networks}

As was the case for NeuralODEs, the convergence of neural networks to a gradient flow has not yet been proved for anything as complex as AI weather models. But there have been some highly suggestive results recently for attention-only transformers, the simple case of a residual network containing only self-attention blocks. The processor block $B_i$ is specified to be an interaction term 
\begin{equation}
    \label{eq:interaction}
    I(\tilde{x}_p) = \sum_q A_{p,q} V \tilde{x}_q,
\end{equation} 
which describes the effect of the state at position $q$ on a target position $p$. The attention is
\begin{equation}
    \label{eq:attention }
    A_{p,q} = \frac{e^{\langle Q\tilde{x}_p, K\tilde{x}_q \rangle}}{\sum_r e^{\langle Q\tilde{x}_p, K\tilde{x}_r \rangle}}.
\end{equation} 
$K$, $Q$ and $V$ are the key, query and value matrices, respectively. \citet{sanderSinkformersTransformersDoubly2022} show that the mean-field limit of this model satisfies a continuity equation, although the velocity $v$ is not a Wasserstein gradient flow since the attention is not symmetric in $\tilde{x}_p$ and $\tilde{x}_q$. 

Other versions of self-attention, which do feature a symmetric self-attention (symmetric $Q^T K$ matrix) can be shown to give gradient flow in a modified Wasserstein metric \citep{geshkovskiMathematicalPerspectiveTransformers2023}.
The resulting flow produces clustering or repulsion of the particles in the latent space, depending on the sign of the interaction term.
But it is unlikely that a symmetric interaction will suffice for a weather model, since the influence of one particle on another is not symmetric. For example, a particle upstream of another one in physical space will influence it through advection by the physical wind, but the downstream particle will not have an equal influence on the upstream one. The interaction term in the network will have an asymmetric $Q^T K$ matrix, or in the case of a graph neural network, a directed graph with two edges between each pair of connected particles, one in each direction. With an asymmetric interaction, the particles will not necessarily cluster or repel each other in the latent space.

Work on neuralODEs and attention only transformers considers the two parts of processor block separately, but one recent study looks at the combination \citep{alvarez-lopezPerceptronsLocalizationAttentions2026}.
Again symmetry of processor operations is assumed in order to produce a gradient flow. In particular, the attention is assumed to be symmetric between particles, and the MLP weights are assumed to be symmetric. 
An idealization of a layer-norm is also introduced, in the form of a projection of the particle positions onto a unit hypersphere. The result is an irrotational flow, that can be written as the gradient of a potential. The flow includes an attraction or repulsion between particles, modified by the MLP contribution. The MLP divides the latent space by hyperplanes, giving subregions on the surface of the hypersphere with a locally quadratic potential. The particles flow towards local minima of the piecewise quadratic potential.

In summary, the current mathematical results are not sufficiently general to claim that AI weather models must satisfy the gradient flow hypothesis, but do provide some useful intuition about how the flow is determined by the learned parameters.
The particle flow is driven by interactions and an external potential corresponding to the MLP layers. The interaction flow is not symmetric; in advection, the upstream particle attracts the downstream one in the latent space by transferring its properties to it, but there is no equal and opposite attraction in the other direction. Instead the upstream particle is attracted toward particles that are upstream of it. Likewise, the MLP influence is not necessarily a gradient flow. The flow might have a rotational component, although it is also possible that this is discouraged by regularization during training, since the distance traveled by the particles would not be minimal. 
In this case the dynamics would be a flow of particles in the latent space towards local minima defined by the MLP weights, with the interaction providing a drift that coordinates particle movements.

\section{Discussion}
\label{sec:discussion}
\subsection{From gradient flow to continuous time evolution}

To complete the picture of what differential equations might be solved by an AI weather model, we must also consider that the model is applied autoregressively, so that each application of the model advances the forecast by finite timestep $\Delta t$. This is often chosen to be 6~hours for global models. The differential equations in time are obtained in the limit as the time step $\Delta t$ goes to zero. A particularly simple outcome for a gradient flow within each timestep would occur if the flow is not only directed toward a minimum of a free energy, but reaches it by the end of the processor layers at each timestep. In the limit of small timestep, the flow then remains close to a minimum state. This is analogous to a reversible process in thermodynamics, that remains in an equilibrium characterized by the minimum of a free energy.
This is an attractive picture, because during processing the flow is influenced by peculiarities of the architecture and training of the network, but it doesn't matter as long as the final minimum is directly constrained by the loss function in training. Different AI weather models will find the same minima. There is no requirement that the flow must reach a minimum, so this must be regarded as an additional hypothesis that must be tested empirically.

\subsection{Extensions to other architectures}

We also briefly consider how the gradient flow picture could be generalized to three categories of AI weather model that generate forecasts in different ways. In the first category, exemplified by FourCastNet \citep{bonevFourCastNet3Geometric2025,pathakFourCastNetGlobalDatadriven2022}, the particles moving through the latent space correspond to Fourier components rather than spatial mesh points. In this case, we would expect that the dynamics would have a similar form to a model based on a spatial mesh, but, with the latent space variables having a different interpretation, it might be difficult to determine if it represents the same underlying equations. On the other hand, if the underlying equations for one model can be identified, it might be possible to transform them between Fourier and grid-point space and show that the two discretizations are equivalent. 

A second category of model, for example AIFS-CRPS \citep{langAIFSCRPSEnsembleForecasting2024}, uses a noise field as input to generate individual ensemble members, and a CRPS loss function to train the model to produce accurate probabilistic forecasts from the ensemble. In the case of AIFS-CRPS, a single noise field is generated at inference time for each ensemble member and introduced into the processing through a conditional layer-norm. The processing for each member is then deterministic and follows the overall architecture described here, so that the gradient flow picture should apply without modification.

Finally, we consider diffusion or flow-matching models that start with a random noise field and evolve it into the forecast in a series of denoising steps. The most common design is a reversed normalizing flow using score-matching, used for example by GenCast \citep{priceGenCastDiffusionbasedEnsemble2024}, ArchesGen \citep{couaironArchesWeatherGenSkillfulComputeefficient2026} and FGN \citep{aletSkillfulJointProbabilistic2025}. This is a particularly simple form that can be shown rigorously to be a Wasserstein Gradient Flow with free energy given by the learned score \citep{peyreMathematicsArtificialIntelligence2025}. Note however that this free energy may not fit the particular form given in Eq.~(\ref{eq:flow-potential}): if the denoising process is stochastic, the free energy will include an entropy term  \citep{peyreOptimalDiffusionTransports2025}. Each step of the denoising uses a transformer-based network, comparable to the deterministic models discussed in this paper. This implies that the underlying equations still describe a gradient flow, but the implementation in a generative diffusion model uses three nested loops to progress in time: an outer loop in time $t$, a middle loop in noise level $\sigma$, and an inner loop in processor layer depth $s$. It will be an interesting challenge to work out the details of such a model.

\subsection{Evidence for gradient flow}
\label{sec:tests}

Although the gradient flow hypothesis cannot be derived rigorously from the structure of the models, we can look for evidence that it describes their observed behavior while computing a forecast. 
One indication of a flow of the latent space coordinates towards stationary state would be that the velocity, i.e. the size of the increments added by each processor layer, decreases as the final state is approached. Unfortunately, it is not possible to directly compare the magnitudes across layers because the latent space vector is only defined up to an arbitrary affine transform that may be different for each layer \citep{elhageMathematicalFrameworkTransformer2021}. In this subsection, we will address this problem in two stages.

To avoid comparing the magnitude of the increments across different layers where the representations my differ by an affine transform, we consider the spatial variance spectra of the latent space channels, which is independent of the absolute magnitude. The motivation for this is the fact that  the largest variability in the atmosphere occurs on large spatial scales, it would be expected that large increments to the state should be correlated over many mesh points. In this case, a tendency for the magnitude of the increments to decrease with processor depth would be associated with decrease in the spatial scale of the increments. Although the absolute magnitude of the latent space vector is arbitrary, the relative size between mesh points at the same layer depth is meaningful. This allows us to compute a characteristic wavenumber of the increment, defined by the spectral centroid: the variance-weighted average wavenumber.

A second issue to be overcome is that we would like to average the spectra over the different latent space channels to get a spectrum that is representative of the increment as a whole. But again this will not be comparable between layers since the component of the latent space vector may be rescaled relative to one another. To correct for this we attempt to align the representation at different layers to a common basis using translators. 
Translators are learned affine transformations that map intermediate latent states $\tilde{X}_i$ to the final latent state $\tilde{X}_F$ \citep{belroseElicitingLatentPredictions2023}. A translator consists of a single learned linear transformation $W_i$ and a single learned bias term $b_i$, which are applied uniformly to latent state vectors at all mesh nodes. When trained and applied across all intermediate states of a model, these translators align intermediate latent states with the structure of the final representation space, improving their comparability across depth. Further details are provided in the appendix. 

\begin{figure}[h]
\centerline{\includegraphics[width=19pc]{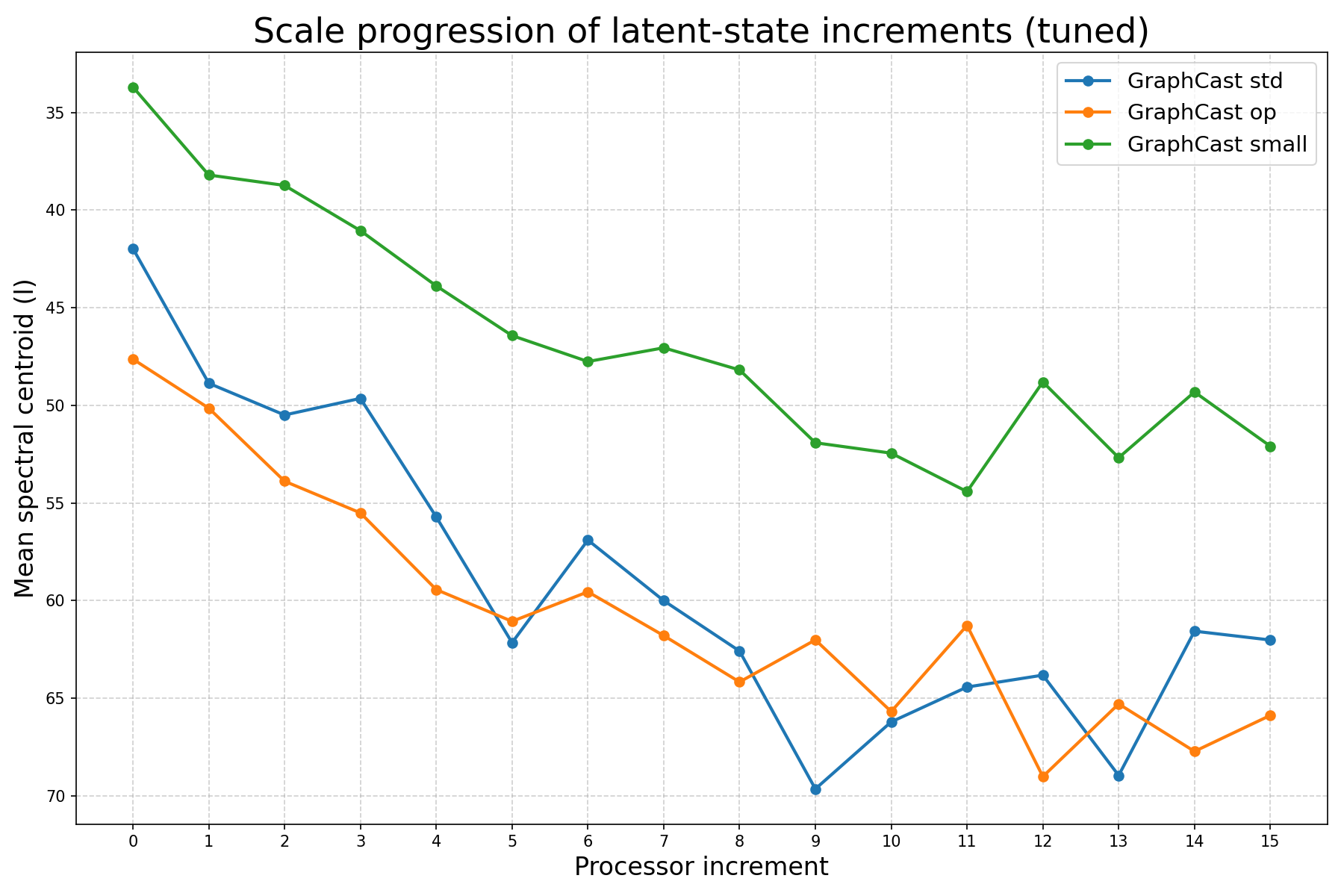}}
 \caption{Spectral centroid (variance-weighted average wavenumber) of variance spectrum averaged over all latent space variables, as function of processor block. Curves show three versions of GraphCast.
}\label{f5}
\end{figure}

\begin{figure}[h]
\centerline{\includegraphics[width=19pc]{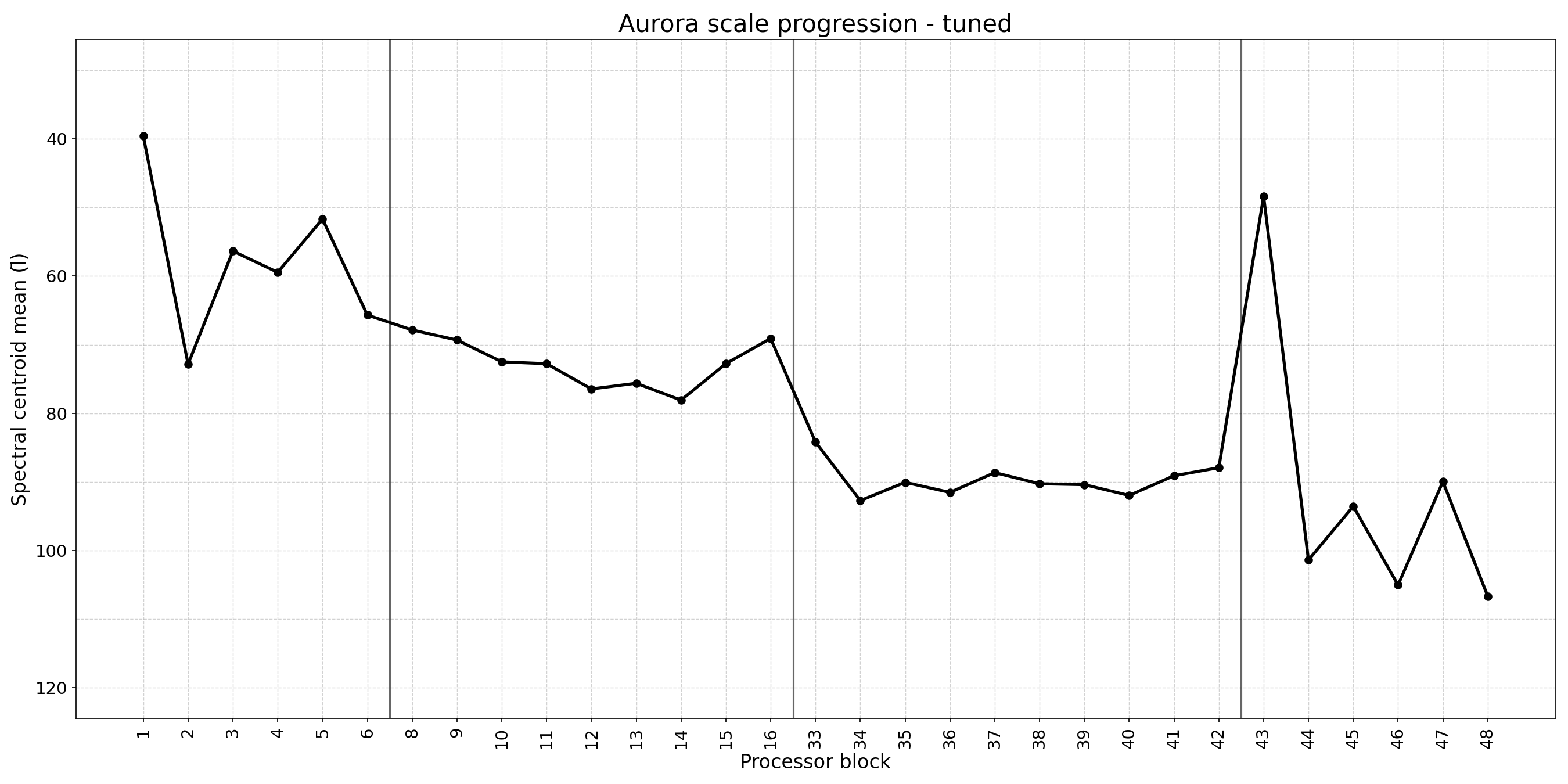}}
 \caption{Spectral centroid (variance-weighted average wavenumber) of variance spectrum averaged over all latent space variables, as function of processor block for the Aurora model. Vertical lines show where the model changes spatial resolution, with higher resolution in Blocks 1-6 and 43-48 than in Blocks 7-16 and 33-42. Note that processor layers 17-32, which correspond to the coarsest resolution grid in the U-net architecture, are omitted because translators have not been computed for these layers. 
}\label{f5a}
\end{figure}

The spatial scale of the latent space increments as a function of processor depth has been computed for the GraphCast  \citep{lamLearningSkillfulMediumrange2023} and Aurora models by computing the spectral centroid, the variance-weighted average wavenumber of the increments for each latent space channel, and then averaging over channels. 
Figure~\ref{f5} plots the scale of the processor increments as function of depth for the three publicly-released versions of GraphCast. Even in the early layers, the spectral centroid is below wavenumber 20, which is approximately the scale of 6h changes in meteorological fields such as wind and temperature, and thus corresponds to the largest scales that the model is trying to predict. The spatial scale of the increments decreases with depth, with the wavenumber of the spectral centroid increasing y a factor of 1.5 by layer 9. After this the scale of the increments is approximately constant. For GraphCast-small, which has lower spatial resolution than the other two versions, the increments are shifted to a slightly larger scale at all processor depths. Fig.~\ref{f5a} shows the corresponding plot for Aurora. A similar trend toward higher wavenumbers is also seen in this model.

\section{Outlook}
\label{sec:outlook}

In this paper we consider the possibility that AI weather models are in fact physical models. 
That is, that they can be regarded as different discretizations of a common underlying set of partial differential equations.
If this hypothesis is true, we have argued that these equations are likely to take a particular form. The latent space vector at each mesh point identifies the position of a particle in the latent space, and in the continuum limit the equations describe the flow of particle density through the latent space. Note that the particles correspond to mesh points that do not move in physical space, and not to fluid parcels in a conventional model. The latent space flow satisfies a continuity equation, with a velocity determined by an interaction term and a position-dependent velocity corresponding to the two parts of a processor block. The key assumption underlying this model is the permutation equivariance of particles, which would hold if the various architectural choices for defining mesh node interactions (graph edge updates, self-attention etc.) can be regarded as numerical approximations to a fully connected system.

To close the set of equations, we propose the Gradient Flow Hypothesis, which suggests that the velocity takes the form of a gradient flow in latent space that drives the particles towards the minimum of a free energy. Spatial truncation of the differential equations corresponds to sampling the density with discrete Lagrangian particles. Temporal truncation creates perturbations away from equilibrium that are reduced during the flow through processor layers.
This view of AI weather models is an unproven hypothesis, but we show preliminary results for two lines of evidence that are at least consistent with the view presented here. Correlations of forecast skill and Centered Kernel Alignment show that AI models are similar to each other in behavior and in internal representation, despite differences in architecture and capacity. Variance spectra of latent space variables show processor layers act on successively smaller spatial scales, consistent with gradient flow towards a stationary state.

This analysis will be continued in a companion paper (Beylich et al., in prep.) that leverages the translators that were briefly described here to perform intermediate decoding: directing the latent space vector at different processor layers to be decoded back to meteorological variables. One result of this study, which is particularly relevant to the gradient flow picture is an analysis of the processor increments for individual variables such as geopotential and temperature. Consistent with Section~\ref{sec:application} of the present paper, it is found that early layers affect larger spatial scales, but it is also possible to show that the decreasing spatial scale coincides with decreasing magnitudes of the increments, as expected for gradient flow.

To thoroughly verify the gradient flow picture, we will need to be able to write down the form of the free energy that is encoded by the interaction and MLP blocks of the processor, including the physical meaning of the latent space variables that it is a function of. Exploring machine learning models in this way is the province of mechanistic interpretability, and a wide variety of techniques are available \citep{bereskaMechanisticInterpretabilityAI2024,raiPracticalReviewMechanistic2025,linSurveyMechanisticInterpretability2025}. To date, there are only a few applications to AI weather models. One example is \citet{tempestMechanisticInterpretabilityTool2026}, who use principal component analysis to identify important directions in the latent space, which can then be visualized and interpreted. A second example \citep{macmillanMechanisticUnderstandingDatadriven2025} uses a combination of sparse autoencoders and probing to identify directions that correspond to meteorological phenomena such as tropical cyclones and atmospheric rivers. 

There is still a long way to go to establish a physical basis for AI weather models. But it is a worthy goal. Knowing a set of equations that the machine learning model is approximating, could lead to new ways to evaluate design decisions for architecture, hyperparameters and training based on how closely the model conforms to the behavior of the underlying equations. Furthermore, it could lead to new ways to judge the physical consistency of models and give reassurance that they will respond realistically when applied to situations not seen in the training data.

\appendix

\section{RMSE correlations}
\label{app:corr}

\begin{figure}[h] 
\centerline{\includegraphics[width=19pc]{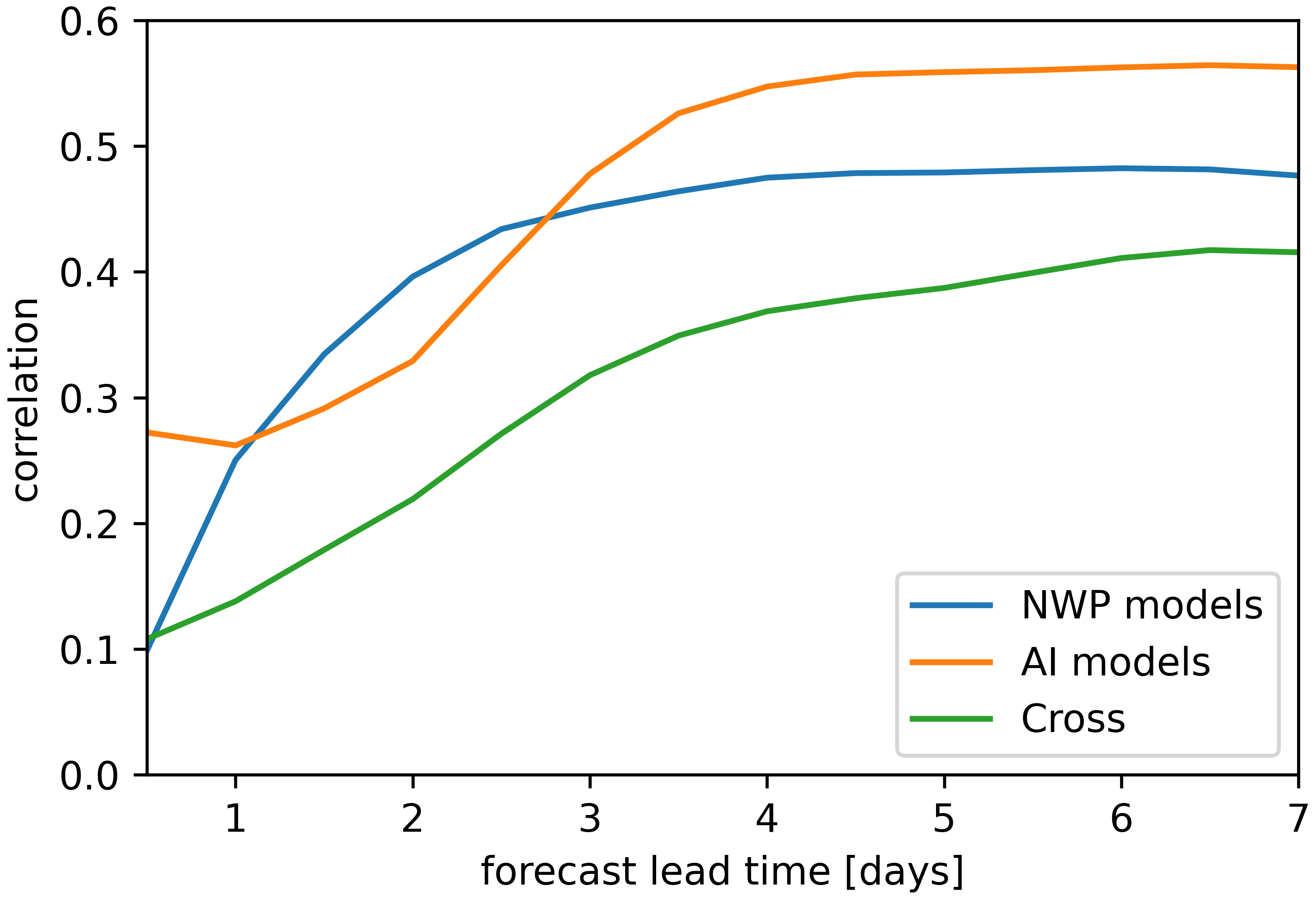}}
 \caption{\label{corrflt}RMSE timeseries correlation over lead time. RMSE is defined as in Fig.~\ref{fcorrmatrix} but calculated for several forecast lead times (12~h to 7~days). The plot shows the average correlation among the NWP model, the AI models and across, respectively (the average over the blocks in Fig.~\ref{fcorrmatrix}).}
\end{figure}

In Fig.~\ref{fcorrmatrix} we showed that the AI model forecasts are more similar among themselves than in comparison to conventional NWP and that this indicates that they have something in common.
There is however one important caveat for interpreting the correlations this way:
The AI models are all started from the same ERA5 analyses, whereas the weather centers are operating different NWP systems, which have their own initial conditions.
Hence the thing they have in common might just be the initial conditions.
It is significant though, that the correlations increase with lead time, reaching a maximum at 5--7 days.
This is illustrated in Fig.~\ref{corrflt}, which shows the average correlation among the AI and NWP models, respectively, and across each other over lead time.
The correlations among the NWP models and the cross correlations also increase with lead time, with the cross correlations always being the lowest.
This behavior makes it unlikely that the common initial condition of the AI models provides a sufficient explanation for the day~5 correlations observed in Fig.~\ref{fcorrmatrix}, and reinforces our hypothesis that they have something else in common which is related to their design.

\section{CKA and CKNNA calculations}
\label{app:cka}

\subsection{Centered Kernel Alignment (CKA)}

The CKA is a metric that measures alignment between the latent spaces of AI models by comparing the similarity between their individual kernels. We use the implementation and code of \cite{huhPlatonicRepresentationHypothesis2024}, the biased version of which is summarized in the following. 

Let \(f\) and \(g\) be representations of two different AI models, that maps inputs, \(x\) and \(y\), into a vector of latent space features, otherwise known as embeddings, of \(\phi_i = f(x_i) = \tilde{X}_i\) and \(\psi_i = g(y_i) = \tilde{Y}_i\), respectively. Here \(\phi_i \in \mathbb{R}^{d_f}\) and \(\psi_i \in \mathbb{R}^{d_g}\), where \(d_f\) and \(d_g\) are their latent dimensions. Note that \(d_f\) does not need to equal \(d_g\).

The embeddings of the final processor step at one mesh node of a model are collected for a range of \(N\) forecast times, such that for representation \(f\), 

\[
\Phi = \{\phi_1, \phi_2, \ldots, \phi_N\}
= 
\begin{bmatrix}
\text{---}\, \tilde{X}_1^{\top} \, \text{---} \\
\text{---}\, \tilde{X}_2^{\top} \, \text{---} \\
\vdots \\
\text{---}\, \tilde{X}_N^{\top} \, \text{---}
\end{bmatrix},
\]

where \(\Phi \in \mathbb{R}^{N \times d_f}\), and all embeddings have been normalized.

Centered kernel matrices using a linear kernel are then created from \(\Phi\) and \(\Psi\):

\[
\bar{\mathbf{K}}_{ij} = \langle \phi_i, \phi_j \rangle - \mathbb{E}_l \big[ \langle \phi_i, \phi_l \rangle \big]
\quad
\bar{\mathbf{L}}_{ij} = \langle \psi_i, \psi_j \rangle - \mathbb{E}_l \big[ \langle \psi_i, \psi_l \rangle \big].
\]
Next the Hilbert-Schmidt Independence Criterion (HSIC) is estimated using the cross covariance of \(\bar{\mathbf{K}}\) and \(\bar{\mathbf{L}}\):

\[
\mathrm{HSIC}(\mathbf{K}, \mathbf{L}) = \frac{1}{(N - 1)^2} \, \mathrm{Trace}(\bar{\mathbf{K}}\bar{\mathbf{L}}).
\]

Normalizing the HSIC provides the CKA: 

\[
\mathrm{CKA}(\mathbf{K}, \mathbf{L}) =
\frac{\mathrm{HSIC}(\mathbf{K}, \mathbf{L})}
{\sqrt{\mathrm{HSIC}(\mathbf{K}, \mathbf{K}) \, \mathrm{HSIC}(\mathbf{L}, \mathbf{L})}}
\]

\subsection{Centered Kernel Nearest Neighbor Alignment (CKNNA)}

To calculate the CKNNA, \(k\) nearest neighbors within each row of the kernel matrices, \(\mathbf{K}\) and \(\mathbf{L}\), are selected. Therefore, at \(k=N\), \(\text{CKNNA}=\text{CKA}\). To select the nearest neighbors, \(\alpha\) is used here:

\[
\mathrm{Align}(\mathbf{K}, \mathbf{L}) = \sum_{i,j} \alpha(i,j)\, \bar{\mathbf{K}}_{ij}\,\bar{\mathbf{L}}_{ij},
\]

where \(\alpha(i,j)\) is equal to \(1\) only if \(\phi_j\) is a nearest neighbor of \(\phi_i\), and if  \(\psi_j\) is a nearest neighbor of \(\psi_i\). It is zero otherwise. The CKNNA can then be calculated by

\[
\mathrm{CKNNA}(\mathbf{K}, \mathbf{L}) =
\frac{\mathrm{Align}(\mathbf{K}, \mathbf{L})}
{\sqrt{\mathrm{Align}(\mathbf{K}, \mathbf{K}) \, \mathrm{Align}(\mathbf{L}, \mathbf{L})}}.
\]

\subsection{Calculation}

Both the CKA and CKNNA are bounded by 0 and 1, whereby 0 shows no alignment and 1 is complete alignment. For our calculations, we use data covering the entire year of 2020, such that \(N=1462\). The two models compared are Graphcast (including the three configurations thereof) and Aurora. GraphCast's latent dimension is \(512\), whereas Aurora's is \(1024\). In the latter case, there are multiple latent levels (4) at each node. To account for this, we concatenated the four latent feature vectors at the relevant mesh node and used this in our calculation.

Seven grid points from across the globe were selected for calculating the metrics, but only minor differences between grid points were found. [add plots from different node locations?]

\begin{figure}[h]
\centerline{\includegraphics[width=19pc]{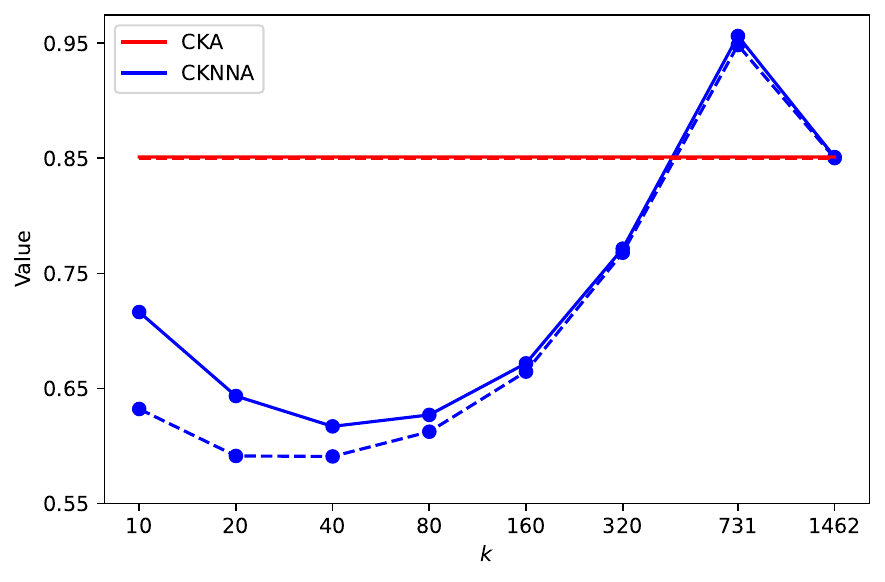}}
 \caption{CKA and CKNNA values for alignment between Aurora and GraphCast Paper models, at various k. Solid (dashed) lines are biased (unbiased) values. Shown for a midlatitude point (50°N, 9°W). (Log x axis).}\label{f7}
\end{figure}

Unbiased counterparts of the CKA and CKNNA are also available (more details in \cite{huhPlatonicRepresentationHypothesis2024}) and were calculated in addition to the biased metrics (Fig.~\ref{f7}). Only small differences were seen, particularly for the CKA. Furthermore, it is seen that as \(k\) varies, the value of CKNNA does too. For our results, we chose \(k\) where the minimum is often, at a value of 40.

\begin{figure}[h]
\centerline{\includegraphics[width=19pc]{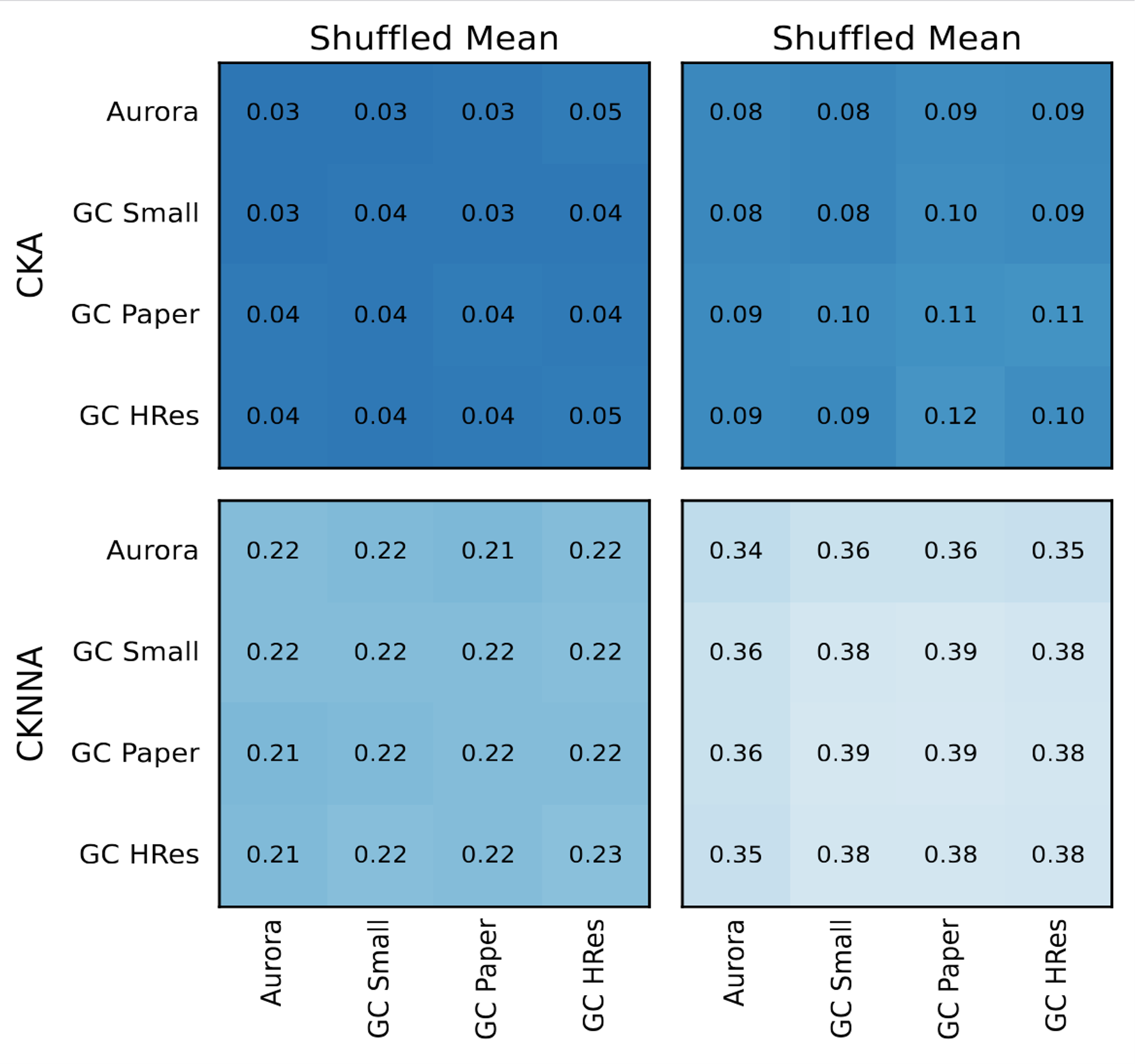}}
 \caption{Block shuffled Centered Kernel Alignment (top row) and Centered Kernel Nearest Neighbor Alignment (bottom row) between Aurora and three versions of GraphCast. First column for a midlatitude point (50°N, 9°W), and second column for a tropical point (1°S, 0°E).}\label{f8}
\end{figure}

A test of significance is provided by block shuffling  \(\Phi\) for one of the models such that the examples are not aligned with \(\Psi\), and then calculating the relevant metric again. A block size of 20, corresponding to a time interval of 5~days is used to avoid serial correlation in the data. This process is repeated 5 times and the results are averaged to obtain the shuffled mean. As can be seen by comparing Fig.~\ref{f8} to Fig.~\ref{f6}, the shuffled CKA and CKNNA values are always substantially below the values obtained after shuffling the data, indicating that the observed correlations are unlikely to have occurred by chance.  

\section{Translators}

Translators were introduced to account for the lack of representational stability in residual networks. Two issues were identified: with the addition of processor increments, the residual stream is not centered at zero but instead accumulates a bias. Second, deep models do not possess a privileged basis, and individual layers may apply arbitrary rotations in latent space that are only undone before decoding, thereby changing the meaning of latent dimensions throughout model depth. Representational drift can be quantified by comparing covariance matrices between latent states, as done by \citet{belroseElicitingLatentPredictions2023} for LLMs. 

To correct for this representational drift and make latent states comparable, we train and employ translators as affine transformations, that align intermediate latent states with the final latent state of a model. A translator is a single affine transformation: 

\begin{equation}
\tilde{X}_F = W_i \tilde{X}_i + b_i,
\end{equation}

where $W_i$ is a learned linear transformation (interpreted as a change-of-basis matrix) and $b_i$ is a learned bias term. This mapping transforms intermediate latent states $\tilde{X}_i$ into the final latent state space $\tilde{X}_F$. 

Translators are applied prior to the computation of power spectra and spectral centroids for all versions of GraphCast and across all latent states. For Aurora, latent states at the coarsest internal mesh resolution (latent states 17--32) do not have an associated translator. Furthermore, latent states 7--16 and 33--41 are aligned with latent representation 42, which is the final state before the upscaling block. 

For training, we used 1462 extracted latent states from 2020 for both GraphCast and Aurora, split into 1162 training samples and 290 validation samples. Training minimizes the mean squared error (MSE) between an intermediate state $\tilde{X}_i$ and the final latent state $\tilde{X}_F$ using the Adam optimizer with a learning rate of $5 \times 10^{-4}$. We use a batch size of 2, where each batch consists of full global mesh node fields; thus, before each weight update, two global mesh node fields are processed. This results in $581$ weight updates per epoch. The weight matrices are initialized as the identity. We observe fast convergence: translators are largely trained after 1--2 epochs and typically converge within approximately 5 epochs. Translators are location-independent; the same matrix and bias term are applied to all spatial locations.

%%%%%%%%%%%%%%%%%%%%%%

% \newpage
% ---------------------------------------------------------------------------------

% \section{Citations}
% Citations to standard references in text should consist of the name of the
% author and the year of publication, for example, \citet{Becker+Schmitz2003} or
% \citep{Becker+Schmitz2003} using the appropriate $\backslash$citet\ or
% $\backslash$citep commands, respectively. A variety of citation formats can
% be used with the natbib package; however, the AMS prefers that authors use only the $\backslash$citet\ and
% $\backslash$citep commands. References should be entered in the references.bib file. For a thorough
% discussion of how to enter references into the references.bib database file
% following AMS style, please refer to the \textbf{AMS\_RefsV5.pdf} document
% included in this package.

%%%%%%%%%%%%%%%%%%%%%%%%%%%%%%%%%%%%%%%%%%%%%%%%%%%%%%%%%%%%%%%%%%%%%
% REFERENCES
%%%%%%%%%%%%%%%%%%%%%%%%%%%%%%%%%%%%%%%%%%%%%%%%%%%%%%%%%%%%%%%%%%%%%
 % This shows how to enter the commands for making a bibliography using
 % BibTeX. It uses references.bib and the ametsoc2014.bst file for the style.

\bibliographystyle{ametsoc2014}
\bibliography{My_Library}

%%%%%%%%%%%%%%%%%%%%%%%%%%%%%%%%%%%%%%%%%%%%%%%%%%%%%%%%%%%%%%%%%%%%%
% FIGURES
%%%%%%%%%%%%%%%%%%%%%%%%%%%%%%%%%%%%%%%%%%%%%%%%%%%%%%%%%%%%%%%%%%%%%
% \begin{figure}[h]
%  \centerline{\includegraphics[width=19pc]{figure01.pdf}}
% % Standard figure sizes are 19 (one column), 27, 33, and 39 (two columns) picas.
%   \caption{Enter the caption for your figure here.  Repeat as
%   necessary for each of your figures. Figure from \protect\cite{Knutti2008}.}\label{f1}
% \end{figure}

\end{document}